\documentclass[aip,amsmath,amssymb,reprint,]{revtex4-2}
\usepackage[colorlinks=true,citecolor=blue,linkcolor=blue,urlcolor=blue]{hyperref}
\usepackage{graphicx}
\usepackage{dcolumn}
\usepackage{bm}
\usepackage[utf8]{inputenc}
\usepackage[T1]{fontenc}
\usepackage{mathptmx}
\usepackage{etoolbox}
\makeatletter
\def\@email#1#2{
 \endgroup
 \patchcmd{\titleblock@produce}
  {\frontmatter@RRAPformat}
  {\frontmatter@RRAPformat{\produce@RRAP{*#1\href{mailto:#2}{#2}}}\frontmatter@RRAPformat}
  {}{}
}
\makeatother
\begin{document}

\preprint{AIP/123-QED}

\title[Single crystal synthesis, structure, and magnetism of Pb$_{10-x}$Cu$_x$(PO$_4$)$_6$O]{Single crystal synthesis, structure, and magnetism of Pb$_{10-x}$Cu$_x$(PO$_4$)$_6$O}
\author{P. Puphal*}
\email{puphal@fkf.mpg.de}
\affiliation{Max Planck Institute for Solid State Research, Heisenbergstra{\ss}e 1, D-70569 Stuttgart, Germany}
\author{M. Y. P. Akbar}
\affiliation{Max Planck Institute for Solid State Research, Heisenbergstra{\ss}e 1, D-70569 Stuttgart, Germany}
\affiliation{Faculty of
Mathematics and Natural Sciences, Institut Teknologi Bandung, Jl. Ganesha 10, 40132 Bandung, Indonesia}
\author{M. Hepting}
\author{E. Goering}
\author{M. Isobe}
\affiliation{Max Planck Institute for Solid State Research, Heisenbergstra{\ss}e 1, D-70569 Stuttgart, Germany}
\author{A. A. Nugroho}
\affiliation{Max Planck Institute for Solid State Research, Heisenbergstra{\ss}e 1, D-70569 Stuttgart, Germany}
\affiliation{Faculty of
Mathematics and Natural Sciences, Institut Teknologi Bandung, Jl. Ganesha 10, 40132 Bandung, Indonesia}
\author{B. Keimer}
\affiliation{Max Planck Institute for Solid State Research, Heisenbergstra{\ss}e 1, D-70569 Stuttgart, Germany}
\date{\today}

\begin{abstract}
The recent claim of superconductivity above room temperature in Pb$_{10-x}$Cu$_x$(PO$_4$)$_6$O with 0.9 < $x$ < 1 (referred to as LK-99) has sparked considerable interest. To minimize the influence of structural defects and impurity phases on the physical properties, we have synthesized phase-pure single crystals with a copper doping level of $x \sim 1$. We find that the crystals are highly insulating and optically transparent. X-ray analysis reveals an uneven distribution of the substituted Cu throughout the sample. Temperature ($T$) dependent magnetic susceptibility measurements for $ 2 \leq T \leq 800$ K reveal the diamagnetic response characteristic of a non-magnetic insulator, as well as a small ferromagnetic component, possibly originating from frustrated exchange interactions in Cu-rich clusters in the Pb$_{10-x}$Cu$_x$(PO$_4$)$_6$O structure. No anomalies indicative of phase transitions are observed. We therefore rule out the presence of superconductivity in Pb$_{9}$Cu(PO$_4$)$_6$O crystals, and provide some considerations on the origin of anomalies previously reported in experiments on polycrystalline specimen.

\end{abstract}
\maketitle

\begin{quotation}

\end{quotation}

\section{\label{sec:level1}Introduction}

Room-temperature superconductivity holds the potential to disrupt various technologies—from enhancing energy storage and grid efficiency to advancing maglev trains, medical imaging, and electronic devices. In this context, recent claims about the presence of superconductivity in powders of the Cu-substituted lead phosphate Pb$_{10-x}$Cu$_x$(PO$_4$)$_6$O with 0.9 < $x$ < 1 (referred to as LK-99) \cite{Lee2023,Lee2023a} have stirred significant anticipation within the condensed matter research community and beyond. In particular, the pronounced resistivity drop and the onset of a large diamagnetic response at a temperatures $T \sim 400$ K have been interpreted as evidence of superconductivity in the initial studies. Yet, efforts to replicate these key observations in polycrystalline specimens have yielded disparate results \cite{Kumar2023,Kumar2023a,liu2023,hou2023,Wu2023,Wu2023a,Guo2023,Shilin2023,Timokhin2023,Jiang2023}, likely due to different levels of impurity phases generated by the complex synthesis via sulfur based compounds. In particular, the initial studies \cite{Lee2023,Lee2023a} reported a significant residue of Cu$_2$S in their powder samples. Moreover, grain boundaries in polycrystalline samples, whose prevalence and properties may depend on the preparation conditions, often exhibit physical properties different from those of the bulk. Conclusive tests of the original claims therefore require a well-defined pure phase, based on single-crystalline samples. 

Motivated by these considerations, we have synthesized phase-pure single crystals of Pb$_{9}$Cu(PO$_4$)$_6$O using the travelling solvent floating zone growth (TSFZ) method, relying solely on phosphate-based compounds as precursors. None of the purported signatures of superconductivity are confirmed in these specimens. In particular, the crystals are highly insulating and optically transparent. Whereas the magnetic susceptibility includes a small ferromagnetic component (likely due to local magnetic order in the inhomogeneous network of Cu substituents), the magnetic susceptibility exhibits no anomalies indicative of phase transitions (including superconductivity) in a wide range of temperatures up to 800 K. We discuss possible origins of the anomalies reported in phase-impure powder samples. In particular, the impurity phase Cu$_2$S exhibits two structural phase transitions, with the one at 398 K coinciding with an insulator-to-metal transition \cite{Abdullaev1968} that may be mistakenly attributed to superconductivity.  

\section{\label{sec:level2}Methods}

A precursor powder was prepared by mixing the corresponding stoichiometries of 9 PbO : 1 CuO and 9 NH$_{4}$H$_{2}$PO$_{4}$. Subsequently, the powder was ball-milled for 20 minutes and the mixture was transferred in alumina crucibles to a box furnace, followed by heating to 750$^{\circ}$C for 10 hours followed by grinding and another heating of 10 hours. 
Cylindrically shaped feed and seed rods were prepared by ball-milling the sintered materials, which were filled into rubber forms with 6~mm diameter. The rubber was evacuated and pressed in a stainless steel form filled with water using a Riken type S1-120 70 kN press. All rods were heat treated at 800$^{\circ}$C. For the first growth, pellets were prepared in a 4mm diameter press form with the Riken press. The single-crystal growth was carried out in a Crystal System Corporation optical image furnace CSC FZ-T-10000. Four halogen lamps operating at 150 W were used as a heating source. 

\begin{figure*}[tb]
\includegraphics[width=2.0\columnwidth]{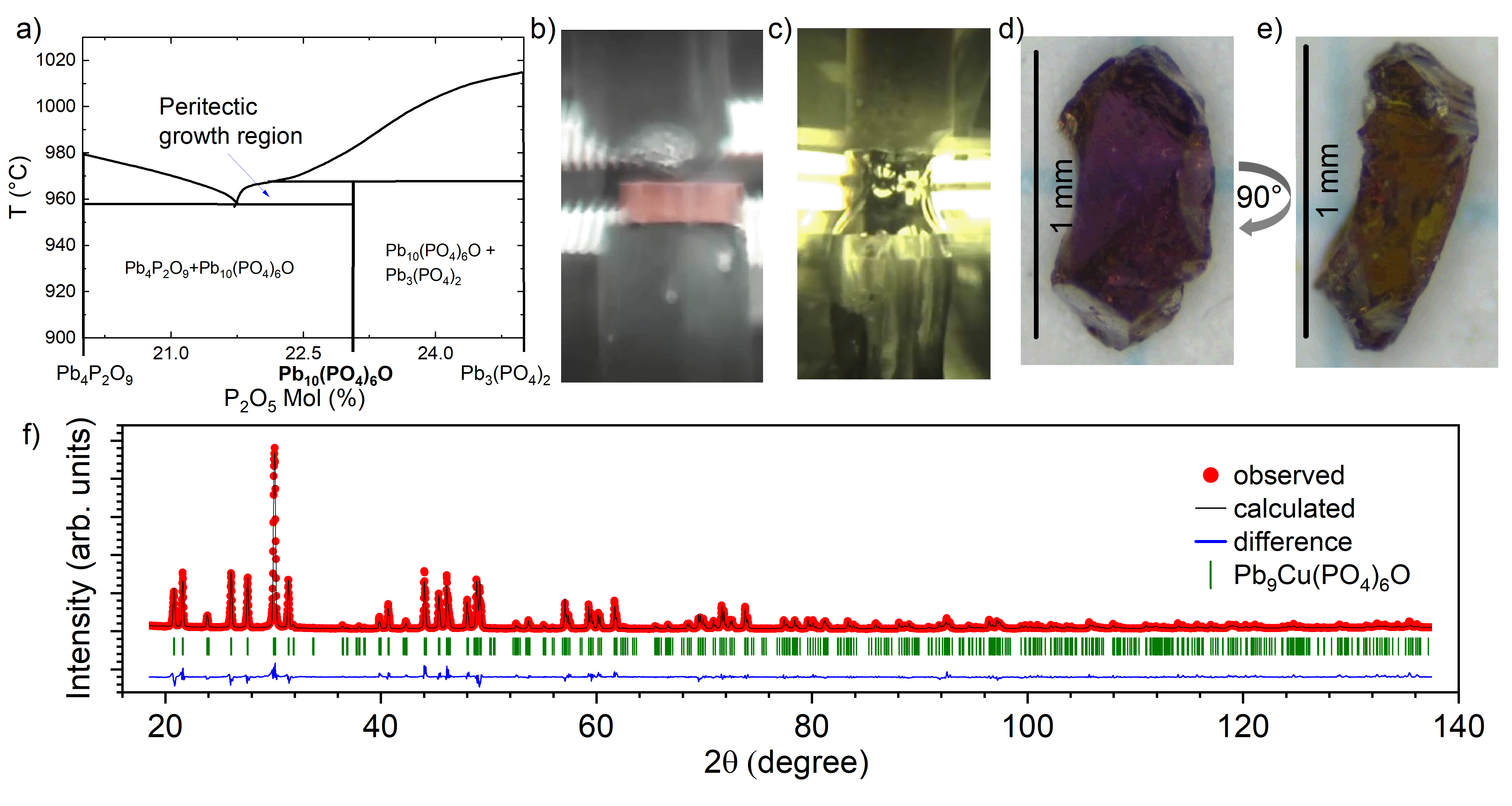}
\caption{a) Selected region of the PbO - P$_2$O$_5$ phase diagram from Ref.~\onlinecite{Merker1960}, with the phase formation temperature plotted as a function of the mol\% of P$_2$O$_5$. The blue arrow indicates the peritectic growth region of Pb$_{10}$(PO$_4$)$_6$O, and vertical thick black lines highlight the three surrounding phases. b) Photograph of the orange PbO flux pellet balancing between the seed and feed rods. c) Photograph of the growth process, utilizing 150 W lamps at 41.6 \% power. d,e) Photographs of a Pb$_{9}$Cu(PO$_4$)$_6$O single crystal under polarized light. Turning the crystal over by 90° leads to a color change from purple (d) to brownish (e), revealing a dichroism effect. f) Powder XRD pattern of a pulverized Pb$_{9}$Cu(PO$_4$)$_6$O single crystal (red data points). The solid black line corresponds to the calculated intensity from the Rietveld refinement, the solid blue line is the difference between the experimental and calculated intensities, and the vertical green bars are the calculated Bragg peak positions. 
}
\label{growth}
\end{figure*}

The as-grown crystals are characterized by optical microscope to select the phases obtained from the growth. The structural properties of each crystals were characterized using powder x-ray diffraction (PXRD) performed at room temperature using a Rigaku Miniflex diffractometer with Bragg Brentano geometry, Cu K$_\alpha$ radiation and a Ni filter. Rietveld refinements were conducted with the FullProf software suite. 
Single crystal diffraction was performed at room temperature using a Rigaku XtaLAB mini II with Mo K$_\alpha$ radiation. The data was analyzed with CrysAlis(Pro) and the final refinement done using Olex2 with SHELX.
Energy-dispersive x-ray spectra  (EDX) were recorded with a NORAN System 7 (NSS212E) detector in a Tescan Vega (TS-5130MM) SEM.
Differential Thermal Analysis studies (DTA) and Thermogravimetry (TG) was done using a Netzsch DTA/TG. For EDX mapping, we used a Thermofischer tabletop Phenom Pharos REM from the MPI-IS, utilizing the high pressure option to prevent charging phenomena.
Magnetic susceptibility measurements were performed using a vibrating sample magnetometer (MPMS 3, Quantum Design). For high temperature measurements, we used the oven option of the MPMS SQUID. Electrical transport measurements were attempted with a Physical Property Measurement System (PPMS, Quantum Design). For EDX mapping, we used a Thermofischer tabletop Phenom Pharos REM, utilizing the high pressure option to prevent charging phenomena. 

\section{\label{sec:level3}Results}

\begin{figure*}[tb]
\includegraphics[width=2.0\columnwidth]{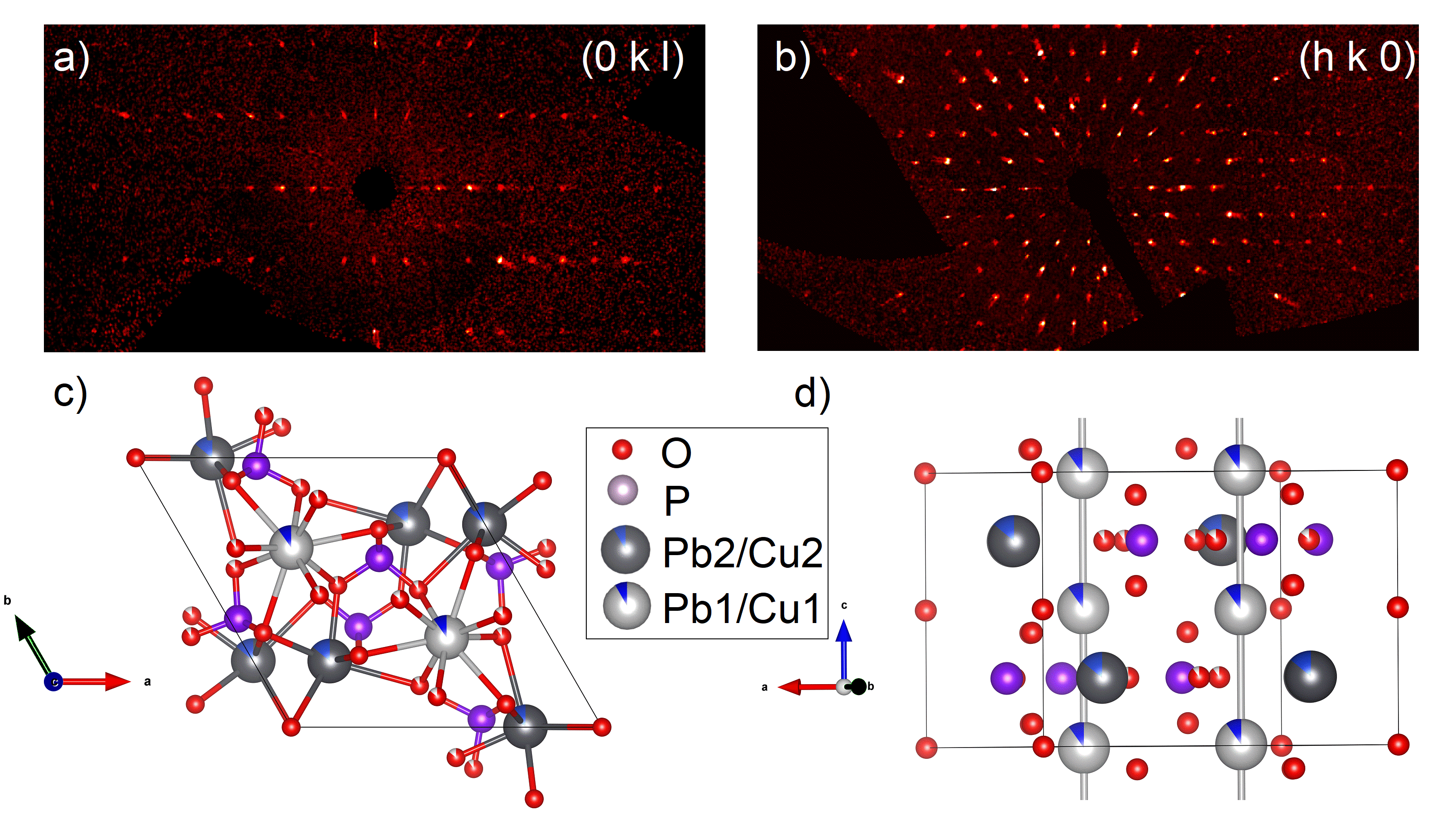}
\caption{(a) Reconstructed XRD intensity map of the (0 k l) plane measured on a single crystal with a size of approximately 1 mm$^3$. The corresponding refinement results are presented in Tab.~\ref{table}. (b) XRD map of the (h k 0) plane. (c) Schematic of the refined crystal structure in $P6_3/m$ in the projection along the c-axis direction, where the colored spheres represent Pb (light and dark grey), P (purple), O (red), and Cu (blue). (d) Crystal structure  in the projection perpendicular to the c-axis direction, with the shortest Pb-Pb bonds highlighted by thick gray lines.
}
\label{diffraction}
\end{figure*}
 
A phase formation diagram for the synthesis of Pb$_{10}$(PO$_4$)$_6$O has already been reported in Ref.~\onlinecite{Merker1960}, and is depicted in Fig.~\ref{growth}a. Using this as a guideline for the growth of Cu-substituted Pb$_{10-x}$Cu$_x$(PO$_4$)$_6$O single crystals, we employ the TSFZ method, where a flux pellet is first placed on the seed rod and then balancing between the seed and feed rods during the growth (Fig.~\ref{growth}b). Our initial growth utilized a pure PbO pellet (batch 1). For the subsequent growths, we prepared pellets with molar PbO:CuO ratios of 9:1 and 1:1 for batch 2 and  3, respectively. The growth chamber was then evacuated and subsequently filled with Ar at 1 bar, maintaining a flow rate of 0.3 l/min. In accordance with the phase formation diagram, we initiated the growth using the PbO self-flux. Upon connecting the rods, a slow growth rate of 1 mm/h was maintained. Due to the decrease of the melting temperature when progressing through the peritectic point, it was necessary to maintain a feed rate of 1 mm/h. This resulted in a broader seed, as illustrated in Fig.~\ref{growth}c. 

Post-growth, the boule was fractured using a mortar, yielding transparent crystals with a purple-brown shade in all batches. A representative crystal of batch 1 is shown in Figs.~\ref{growth}d,e. The change of shading upon rotation of the polarized light propagation direction (dichroism) confirms the single-crystalline nature of our specimens. A powder XRD analysis of a crushed crystal from batch 1 indicated a phase-pure Pb$_{9}$Cu(PO$_4$)$_6$O composition with the hexagonal $P6_3/m$ structure (space group \#176). The refinement result is summarized in Tab..~\ref{table} and indicates a stoichiometry of Pb$_{8.9(1)}$Cu$_{1.1(2)}$P$_6$O$_{25}$.

As a next step, we performed multiple single-crystal XRD measurements, specifically aimed at re-evaluating the details of the crystal structure of the Cu substituted variant, which were previously determined only from powder samples. The result of a single crystal of batch 3, indicate a stoichiometry of Pb$_{8.8(3)}$Cu$_{1.2(3)}$P$_6$O$_{25}$, and are summarized in Tab.~\ref{table}. The reconstructed XRD maps along with the crystal structure are shown in Fig.~\ref{diffraction}. Notably, the refinement of our XRD data allows for a determination of the Cu site occupancy with a higher confidence. Furthermore, the result of the refinement suggests that the Cu substitution coincides with a reduction in the oxygen occupation, likely attributable to the square planar coordination of Cu. Overall, the XRD analysis indicates that the average Cu doping level in our crystals falls into the regime where the initial studies on powder samples \cite{Lee2023,Lee2023a} claimed the occurrence of superconductivity. 
Summarizing our XRD results we have obtained a large varying lattice constants moving in the range of the two presented in Tab.~\ref{table}. As our powder data of one crushed crystal in Fig. \ref{growth} from
batch 1 show  a,b=  9.85319(6) \AA, and c = 7.44154(5) \AA; another refinement of powder data that is not shown on a crushed crystal of batch 3 yields a,b= 9.84847(12) \AA, and c = 7.43811(11) \AA, while single crystal XRD of batch 1 a,b= 9.8380(9)  \AA, and c = 7,422(11) \AA, batch 2 a,b = 9.8140(18) \AA, and c = 7.4370(15) \AA, batch 3 a,b = 9.778(2) \AA, and c = 7.3966(15) \AA and finally the presented one in Fig.~\ref{diffraction} of batch 3 a,b= 9.798(2) \AA, and c =  7.3984(15) \AA.
As discussed further below Cu is inhomogenously distributed, Pb 2+ with a coordination of VIII has a large ionic radius of 1.29, while Cu is not even reported for such a high coordination, hence the oxygen vacancies that we find in refinement. Cu maximally reaches a ionic radius of 0.73 as Cu2+ with a coordination of VI. Thus we expect a lattice contraction with substitution. Thus in accordance with expectations we see a contraction of the lattice with enhanced Cu substiution.

\begin{table}[tb]
\caption{Refined atomic coordinates of Pb$_{9}$Cu(PO$_4$)$_6$O in the hexagonal $P6_3/m$ (\#176) structure, extracted from powder XRD (first table) and single crystal XRD measurements (second table). The obtained lattice constants are a, b = 9.85319(6) \AA, and c = 7.44154(5) \AA~ for the powder data and a, b = 9.7393(19) \AA, and c = 7.3953(12) for the single crystal data.}
\begin{tabular}{lllllll}
Atom & Label & $x$ & $y$ & $z$ & Uiso (\AA$^2$) & Occ. \tabularnewline\hline
Pb & Pb1 & 0.333333 & 0.666667 & 0.0003(10) & 0.0141(6) & 0.89(2) \tabularnewline
Cu & Cu1 & 0.333333 & 0.666667 & 0.0003(10) & 0.0141(6) & 0.11(2) \tabularnewline
Pb & Pb2 & 0.7529(3) & 0.7550(4) & 0.25 & 0.0181(5) & 0.89(2) \tabularnewline
Cu & Cu2 & 0.7529(3) & 0.7550(4) & 0.25 & 0.0181(5) & 0.11(2) \tabularnewline
P & P1 & 0.4093(16) & 0.3821(15) & 0.25 & 0.022(4) & 1 \tabularnewline
O & O1 & 1 & 1 & 0.5 & 0.011(4) & 1 \tabularnewline
O & O2 & 0.365(4) & 0.515(3) & 0.25 & 0.011(4) & 1 \tabularnewline
O & O3 & 0.606(4) & 0.462(3) & 0.25 & 0.011(4) & 0.86(6) \tabularnewline
O & O4 & 0.3579(20) & 0.2580(19) & 0.074(2) & 0.011(4) & 1 \tabularnewline
&&&&&&  \tabularnewline
Atom & Label & $x$ & $y$ & $z$ & Uiso (\AA$^2$) & Occ. \tabularnewline\hline
Pb & Pb1 & 0.333333 & 0.666667 & 0.0040(3) & 0.0040(9) & 0.90(6) \tabularnewline
Cu & Cu1 & 0.333333 & 0.666667 & 0.0040(3) & 0.0040(9) & 0.10(6) \tabularnewline
Pb & Pb2 & 0.7531(2) & 0.7546(3) & 0.25 & 0.0060(8) & 0.86(6) \tabularnewline
Cu & Cu2 & 0.7531(2) & 0.7546(3) & 0.25 & 0.0060(8) & 0.14(6) \tabularnewline
P & P1 & 0.4026(16) & 0.3724(14) & 0.25 & 0.001(3) & 1 \tabularnewline
O & O1 & 1 & 1 & 0.5 & 0.008(19) & 1 \tabularnewline
O & O2 & 0.336(5) & 0.490(4) & 0.25 & 0.002(8) & 0.90(6) \tabularnewline
O & O3 & 0.587(4) & 0.474(4) & 0.25 & 0.007(8) & 0.86(6) \tabularnewline
O & O4 & 0.351(4) & 0.266(4) & 0.084(4) & 0.023(8) & 1 \tabularnewline
\end{tabular}
\label{table}
\end{table}

\begin{figure}[tb]
\includegraphics[width=1.0\columnwidth]{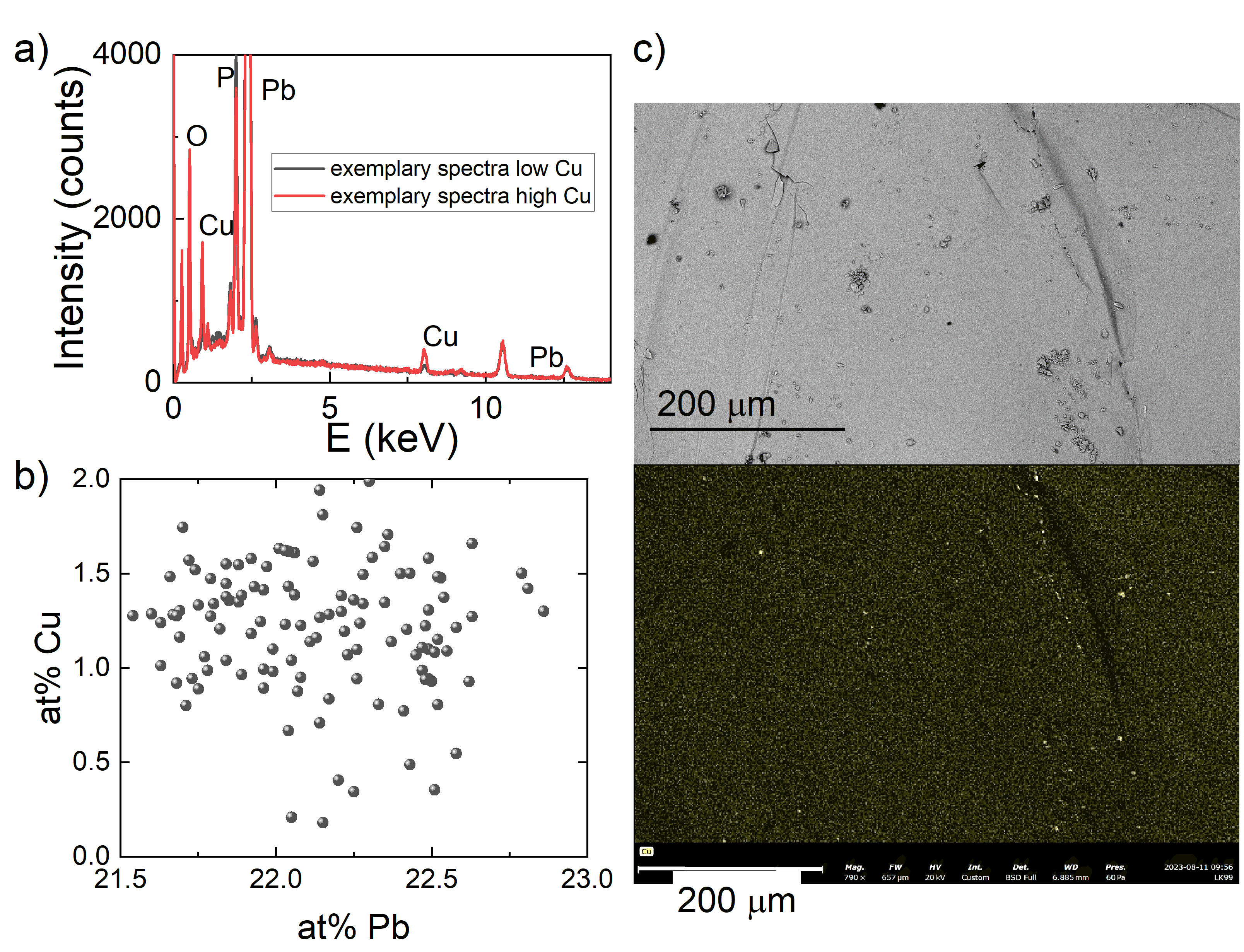}
\caption{a) Selected EDX spectra from a Cu rich (red line) and Cu deficient (black line) region on the cleaved surface of a Pb$_{10-x}$Cu$_x$(PO$_4$)$_6$O single crystal. b) Distribution of the Cu/Pb substitution in at\%, as extracted from EDX spectra taken on various points on the surface of a single crystal. The measurement time for each EDX point was 30 s. c) SEM image (top) and the corresponding EDX map (bottom), where the spatial distribution of the EDX intensity of Cu is highlighted in yellow.
}
\label{EDX}
\end{figure}

Figure~\ref{EDX} summarizes EDX data compiled from over 20 broken pieces of crystals across three  batches. Notably, when examining the crystals in a scanning electron microscope, we observe a charging effect from the electron beam. This suggests that the crystals are highly insulating, which was corroborated by transport measurements (see below). A closer look at two EDX spectra, taken from different locations on a single crystal's surface (Fig.~\ref{EDX}a), reveals a varying Cu content as best seen in the EDX map shown in Fig.~\ref{EDX}c). This implies that the Cu doping is not distributed homogeneously in the crystals. However we highlight that Cu is incorporated in the matrix as it can be found everywhere in the single crystalline boule and is observed in single crystal XRD refinement as shown before. To shed more light on this observation, we plot the distribution of the atomic percentage (at\%) of the Cu substitution for several EDX measurement locations in Fig.~\ref{EDX}b, illustrating a large variation of Cu substitution levels. Due to the strong variation among the EDX points, the error bar for the average Cu and Pb stoichiometry of a crystal is relatively large. The determined average stoichiometries are Pb$_{8(1)}$Cu$_{0.4(0.4)}$P$_6$O$_{25}$ for batch 1 grown with PbO flux, Pb$_{9(1)}$Cu$_{0.6(0.4)}$P$_6$O$_{25}$ for batch 2, and Pb$_{9(1)}$Cu$_{0.6(0.7)}$P$_6$O$_{25}$ for batch 3. 
Note that our single-crystal XRD measured in transmission provides an accurate representation of the substitutional content averaged across the entire crystal.

Regarding the electronic transport properties, the optical transparency of our crystals (Fig. 1 d,e) suggests an insulating or semiconducting nature. However, there have been some reports of superconductivity coexisting with optical transparency either due to an exceptionally low plasma frequency \cite{Soma2020,Ohsawa2020,Batson2023} or due to electronic phase separation into insulating majority and superconducting minority phases \cite{Charnukha2012,Charnukha2012a}. We therefore attempted to perform four-point DC resistivity measurements in a PPMS setup (see Methods), but found resistance values above several M$\Omega$ at room temperature. This behavior is also in line with the charging effects observed in the electron microscopy investigation presented above. Together, these findings conclusively falsify the claim of room-temperature superconductivity in LK-99. \cite{Lee2023,Lee2023a}

\begin{figure}[tb]
\includegraphics[width=1.0\columnwidth]{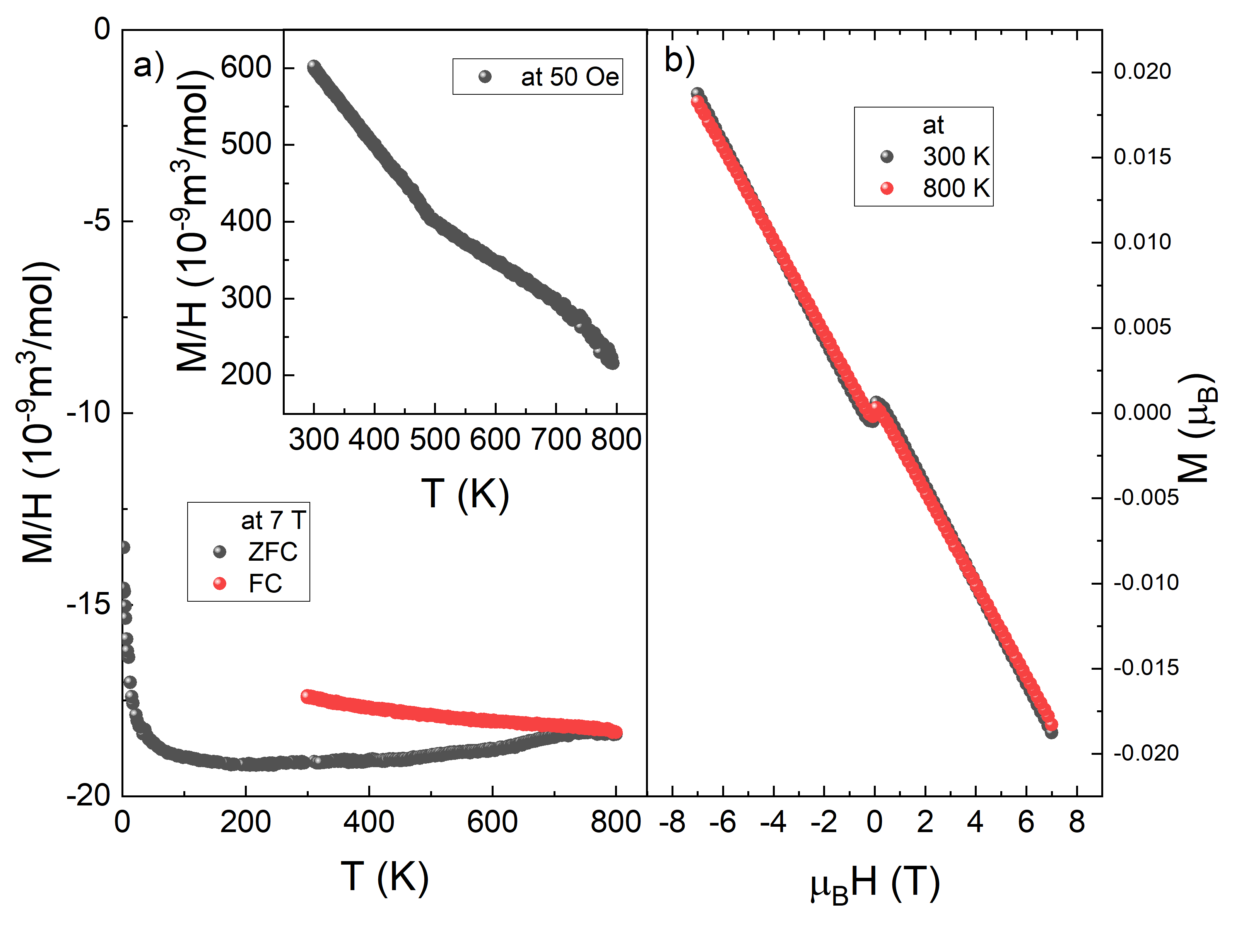}
\caption{a) Temperature dependence of the the magnetic susceptibility in SI units at 7 T measured in field cooled (FC) and zero field cooled (ZFC) mode of a crystal from batch 2 combining low temperature and heater options. The inset shows the magnetic susceptibility in SI units in a field of 0.005 T b) Magnetization versus field measured at 300 and 800 K of the same crystal.
}
\label{Susceptibility}
\end{figure}

We next discuss the magnetic response of the crystals between 2 to 800 K. A first observation is a remarkably low measured signal strength, and hence we note that a small temperature-independent offset might be present in our data, originating from a minute signal from the sample holder, which was not subtracted. For measurements in a small applied field of 5 mT, we detect a positive magnetic susceptibility signal at all temperatures (Fig. \ref{Susceptibility}a inset), but no indications for a clear phase transition besides a tiny kink at 490 K. Certainly there is no transition into a superconducting phase. Figure~\ref{Susceptibility}a) shows the magnetic susceptibility measured in a strong field of 7 T, revealing a diamagnetic response of the crystal at least up to 800 K. The measurement was carried out on a crystal from batch 2. To verify the thermal stability of our crystals and confirm that they do not decompose at elevated temperatures, we conducted a differential thermal analysis (DTA/TG) up to 820 K. Importantly, no mass loss was observed during this analysis (not shown here). In addition, we carried out powder XRD measurements both before and after heating, where we did not observe significant changes compared to the XRD of the Pb$_{9}$Cu(PO$_4$)$_6$O sample shown in Figure \ref{growth}f.

We tried to evaluate the real signal strength of the diamagnetism in this compound by using the room temperature value for the largest sample mass of 63 mg that we measured in a field of 1 T. In this case we carefully measured the background signal and subtract it from our signal. Here, we observed a moment of $-2.5 \cdot 10^{-4}$ emu and a background signal of $-1.46 \cdot 10^{-4}$ emu reulting in $-1.04 \cdot 10^{-4}$ emu for a sample of the volume of 3 x 3 x 1 mm$^3$. Thus our volume susceptibility amounts to $-\frac{1.04\cdot10^{-4}}{1000}\cdot\frac{4\pi\cdot10^{-6}m^{3}}{3\cdot3\cdot1\cdot10^{-9}m^{3}}\approx-1.452\cdot10^{-4}$, which is a value that is remarkably, comparable to Bi (notably for a superconductor we expect -1).

\begin{figure}[tb]
\includegraphics[width=1.0\columnwidth]{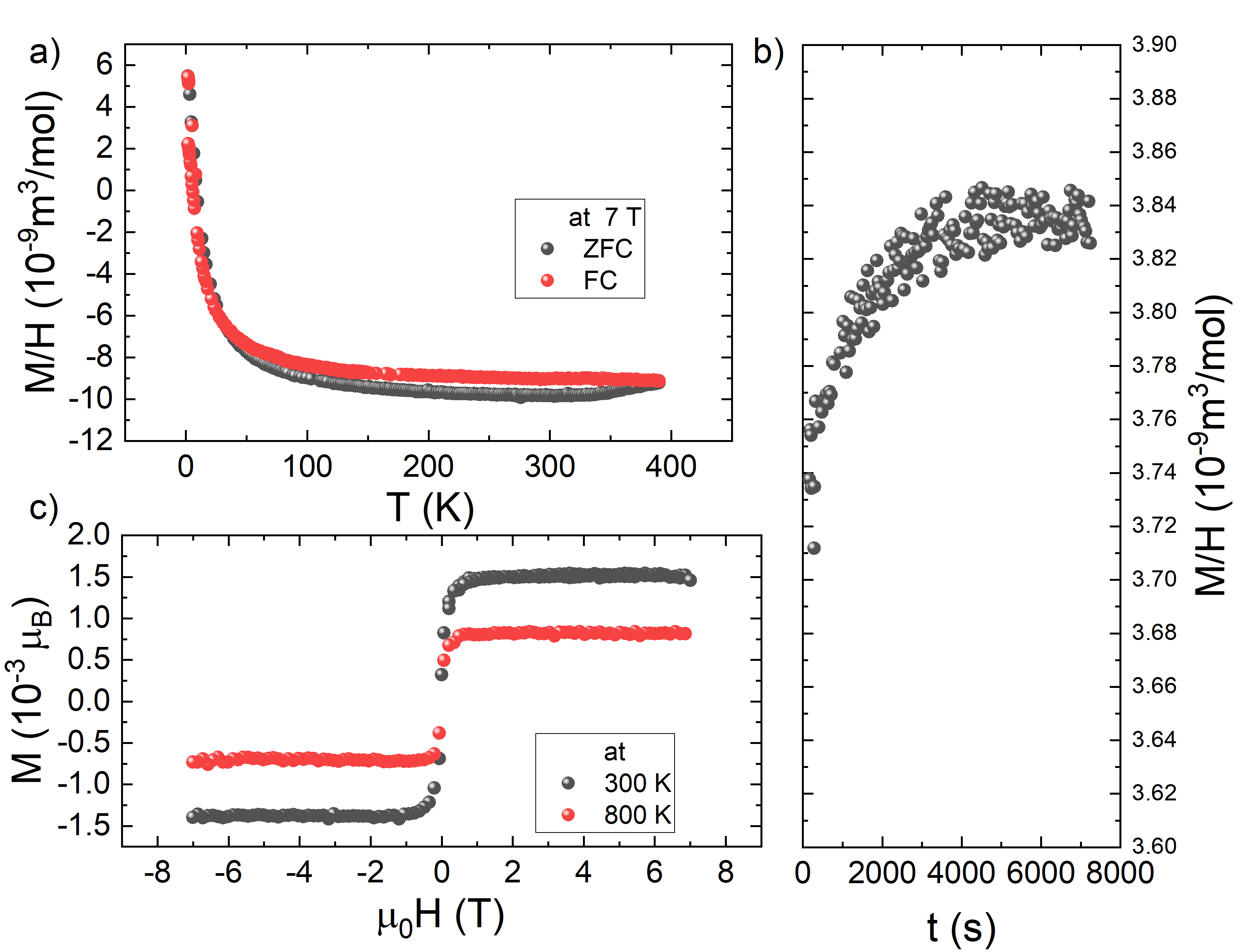}
\caption{a) Temperature dependence of the magnetic susceptibility in SI units of a crystal of batch 3 with the highest Cu content at 7 T in field cooled (FC) and zero field cooled (ZFC) mode only utilizing the standard MPMS 3. b) Time dependence of  the magnetic susceptibility of the same crystal measured at a stable temperature of 2 K under a field of 7 T. c) Magnetization versus field in milli $\mu_B$ at 300 K and 800 K of the data shown in Fig.~\ref{Susceptibility}b) with a linear diamagnetic background subtraction.}
\label{b3}
\end{figure}

For zero field cooling (ZFC) a Curie tail due to isolated paramagnetic impurities is observed at low temperatures, which leads to an upturn of the susceptibility (Fig.~\ref{Susceptibility}a)). Nevertheless, the overall magnetic response remains negative (diamagnetic) down to the lowest measured temperatures of 2 K. For field cooling (FC) from 800 K, we observe a continuous rise in the susceptibility as the temperature decreases. The termination of the FC measurement below 300 K is due to technical constraints of our measurement setup (see Methods). The difference between the ZFC and FC curves persists up to 800 K, where the curves merge due to the here employed measurement protocol heating to a maximum of 800 K. Note that this intersection simply reflects the temperature where the cooling sequence is started, e.g. 390 K for the sequence shown in Figure~\ref{b3}a. This behavior hints at spin glass dynamics, possibly arising from partially ordered magnetic clusters within the sample. (Note that we cannot rigorously rule out the presence of minute amounts of ferromagnetic Fe impurities due to contaminated starting materials, but such contamination is below the detection limits of both EDX and XRD measurements.)
Indeed, a magnetic field sweep at selected temperatures (Fig.~\ref{Susceptibility}b) reveals an initial increase of the sample magnetization for increasing fields, pointing to the presence of ferromagnetic or canted-antiferromagnetic correlations. We ascribe this phenomenon to Cu clusters formed due to the inhomogeneous substitution, resulting in a cluster spin glass effect also proposed theoretically for the material \cite{Si2023}. Importantly, when exposed to stronger magnetic fields, this effect saturates, and the material exhibits a purely diamagnetic response (Fig.~\ref{Susceptibility}b). Further evidence for a cluster spin glass state is derived from the time dependence of the magnetic susceptibility signal, observed for a measurement at 2 K in a constant field of 7 T (Fig.~\ref{b3}b). The glassy dynamics of the magnetic susceptibility, with saturation on a characteristic time scale, is typical for disordered systems. Note that for this measurement the temperature was firmly stabilized at 2 K, hence ruling out potential temperature fluctuations as the origin of the observed time dependence. 

Figure~\ref{b3} displays the magnetic susceptibility of a crystal from batch 3, corresponding to the highest realized Cu concentrations. Notably, for this sample the amplitude of the paramagnetic upturn at low temperature exceeds the diamagnetic response (Fig.~\ref{b3}a). Moreover, we find an enhanced ferromagnetic response for small applied fields (not shown here). Again we highlight the small value of this ferromagnetic part. A possibile scenario is that both the Curie tail and the ferromagnetic (or canted-antiferromagnetic) response are due to the Cu substituents, as expected since this is the only d-electron species in the crystal structure. According to electronic-structure calculations, Cu is divalent and therefore indeed carries a nonzero spin magnetic moment.\cite{Si2023} 

 In Figure~\ref{b3}c we have subtracted the linear diamagnetic contribution, thus highlighting the ferromagnetic contribution versus the applied field. The saturated magnetization amounts to only about 0.001 $\mu_B$ per Cu, and is therefore much smaller than 1 $\mu$B, the full spin-1/2 moment of Cu$^{2+}$. We note, however, that partial Cu substitution in Pb$_{9}$Cu(PO$_4$)$_6$O is expected to generate a complex (and likely highly disordered) network of Cu-O-Cu bonds, which may elicit strong antiferromagnetic exchange interactions (Fig. 2c,d). Canting of the Cu moments due to magnetic frustration and/or local inversion symmetry breaking can then result in canted antiferromagnetism at high temperatures, as observed for instance in nanocrystalline CuO (see Ref. \onlinecite{Batsaikhan2020} and references therein). The connectivity of this network (and hence the amplitude of the net magnetic moment) may depend on the preparation conditions and might be enhanced at grain boundaries in polycrystalline samples. Regardless of its origin, the weakly ferromagnetic response, perhaps in conjunction with the diamagnetic background, may well be at the heart of the partial levitation of Pb$_{9}$Cu(PO$_4$)$_6$O above a permanent magnet at room temperature \cite{Lee2023,Lee2023a,Guo2023,Wu2023}. 

\section{\label{sec:level4}Summary}
In summary, we have successfully grown single crystals of Pb$_{10-x}$Cu$_x$(PO$_4$)$_6$O (LK-99) via the TSFZ method. The phase-pure crystals are highly insulating, suggesting that previously reported transport anomalies in powder samples, such as a insulator-to-metal transition that might have been mistakenly interpreted as a superconducting transition, are likely attributable to Cu$_2$S impurity phases\cite{Zhu2023}. The diamagnetic response of our crystals is broadly consistent with expectations for a non-magnetic insulator, although we also detect weakly ferromagnetic correlations, presumably originating from an inhomogeneous distribution of the Cu substituents. No anomalies indicative of phase transitions are detected for temperatures up to 800 K. These results suggest that the previously claimed occurrence of room-temperature superconductivity in LK-99 is highly unlikely.

\section*{References}

\bibliography{Bib}

\end{document}